\documentclass[doublespacing]{elsart}
\usepackage{natbib}
\usepackage{amssymb}
\begin{document}
\begin{frontmatter}
\title{The distribution of wealth in the presence of altruism for
simple economic models}
\author{M. Rodr\'{\i}guez-Achach\corauthref{qw}}
\ead{achach@gema.mda.cinvestav.mx}
\author{and R. Huerta-Quintanilla}
\address{\it Departamento de F\'{\i}sica Aplicada Unidad M\'erida, 
Cinvestav-IPN,
Km.\ 6 Carretera Antigua a Progreso, M\'erida 97310, Yucat\'an, M\'exico.}
\corauth[qw]{Corresponding author}
\begin{abstract}
We study the effect of altruism in two simple asset exchange
models: the yard sale model (winner gets a random fraction of
the poorer player's wealth) and the theft and fraud model (winner
gets a random fraction of the loser's wealth). We also introduce
in these models the concept of bargaining efficiency, which makes
the poorer trader more aggressive in getting a favorable deal thus
augmenting his winning probabilities. The altruistic behavior
is controlled by varying the number of traders that behave altruistically
and by the degree of altruism that they show. The resulting wealth 
distribution is characterized using the Gini index. We compare
the resulting values of the Gini index at different levels of altruism
in both models. It is found that altruistic behavior does lead to
a more equitable wealth distribution but only for unreasonable
high values of altruism that are difficult to expect in a real
economic system. 
\end{abstract}
\begin{keyword} economic models; econophysics; altruism
\PACS 87.23.Ge\sep 89.65.Gh\sep 89.90.+n
\end{keyword}

\end{frontmatter}

\section{Introduction}

The study of wealth and income distributions in an economical
system is a problem of interest from both the practical and
theoretical points of view and, as expected, has a long history. 
Pareto did some of the first studies on the subject \cite{pareto}.
He proposed that the wealth and income distributions obey universal
power laws, but subsequent studies have shown that this is not
the case for the whole range of wealth values. Mandelbrot
\cite{mandelbrot} proposed that the Pareto conjecture only 
holds at the higher values of wealth and income. The initial part
(low wealth or income) of
the distribution has been recently identified with the Gibbs
distribution \cite{chakra,dragu,chakra2}, while the middle
part, according to Gibrat \cite{gibrat}, takes the form
of a log--normal distribution.

Very recently, this and other aspects of the economy have been
treated under the ``econophysics" point of view, mainly
applying the ideas and tools of statistical mechanics and
Monte Carlo simulations with some degree of success and promising
results (economists however, are still very skeptical about
results obtained from these methods, see \cite{leombruni} for
an interesting discussion). The wealth distribution of any
country, as many other economic quantities, results from very 
complicated processes involving production, taxes, regulations 
and even fraud. Despite this complexity,
very simple models that provide some insight into the whole
process have been devised that qualitatively reproduce
some of the features of real economies. 

We can treat an economy in its simplest form as an interchange
of wealth between pairs of people, or ``agents" at 
successive instants of time (See Hayes \cite{hayes} for an
interesting review). Every time two agents interact,
wealth flows from one to the other according to some rule.
In the so--called ``yard sale" (YS) model, the winner takes a 
random fraction of the wealth of the poorer player, while
in the ``theft and fraud" (TF) model, the winner takes a random
fraction of the loser's wealth. There is no production
or consumption of wealth in these models, nor taxes, savings etc. 
Under these circumstances,
the yard sale model produces a collapse of the economy:
all the wealth ends in the hands of a single agent, 
a phenomenon known as condensation. The theft and fraud
model on the other hand does not collapse but leads to a 
wealth distribution given by the Gibbs distribution. 
See \cite{ispolatov,chakra2,dragu} for details.

The two models mentioned above are oversimplified,
toy--model versions of a real economy, and several
authors have made some refinements to introduce more
realistic situations, for example, allowing the
agents to go into debt \cite{dragu2}, change in the
agents' probability of winning according to the relative
wealth of the traders \cite{sihna}, constant and fractional
savings \cite{chakra3}, and altruistic behavior \cite{trigaux},
among others. In particular,  the introduction of altruism 
in these models has not been
studied in depth, and therefore in this paper we investigate the
effect that altruistic behavior has on the dynamics of the
models and the changes that can produce in the distribution
of wealth.

\section{Models}

In all models we use a fixed number $N$ of individuals with an identical
initial amount of money $m$ to trade. The total wealth of the
community, $Nm$, remains fixed in time. At each time step, two
traders $i$ and $j$ are chosen at random. The winner (which is also
randomly chosen) takes an amount $T$ from the loser. The traders
wealth $w$ at time $t+1$, assuming that $i$ is the winner, will be
\begin{eqnarray}
w_i(t+1)&=&w_i(t) + T\\
w_j(t+1)&=&w_j(t) - T.
\end{eqnarray}
Then, another two traders interact, and the process is repeated $N$ times, 
which constitutes one Monte Carlo step (MCS). The amount $T$ 
of the transaction is defined as
\begin{equation}
T = \alpha MIN(w_i(t),w_j(t)),
\end{equation}
for the YS model and
\begin{equation}
T = \alpha w_j(t),
\end{equation}
for the TF model assuming that agent $j$ loses the transaction.
The parameter $\alpha$ is a uniformly distributed random number in the
interval [0,1].

Altruistic behavior is introduced in the above models
in the following way. First, a certain fraction $p$ of the
N traders is defined as altruists. An altruistic agent remains
in that condition for the whole simulation. Second, a rate
of altruism $r$ is defined and is the same for all of the $pN$
altruistic agents. Suppose that agents $i$ and $j$ trade
at time $t$ and that $i$ is richer than $j$. If agent $i$
wins and is an altruist we will have
\begin{eqnarray}
w_i(t+1) = w_i(t) + T -r(\Delta+T),\\
w_j(t+1) = w_j(t) - T +r(\Delta+T),
\end{eqnarray}
where $\Delta = (w_i-w_j)/2$. With this definition, if
an agent is not altruistic at all ($r=0$), the transactions
proceed as in the original YS and TF models. If the agent
is totally altruistic ($r=1$), the richer agent will give the
other enough of his money so that their fortunes become equal.

Since both of the previous models can be considered too simplistic
to represent an economy, several attempts have been made to
make these models more realistic, as mentioned in the introduction.
Here we follow Sihna \cite{sihna}, who introduces a the concept of
``bargaining efficiency": a rich agent who owns 1000 units
and loses 1 unit during a deal is only losing a 0.1\% of
his wealth. However, an agent who loses the same 1 unit but
whose wealth is of only 5 units is losing 20\% of his fortune.
Therefore it is expected that in a trade between a rich and a
poor agent, the poorer will be more aggressive in getting a 
favorable deal, and that the aggressiveness will be a 
function of the relative wealths of the agents.

The implementation of the above concept is made via the
following ``Fermi function":
The probability that agent $i$ wins in a trade with agent $j$
is given by:
\begin{equation}
p(i|i,j)=\frac{1}{1+\exp(\beta[\frac{x_i}{x_j}-1])},
\end{equation}
where $\beta$ parametrizes the significance of the relative
wealth of the agents. For any $\beta>0$, the poorer agent
has a greater probability of winning the trade. 

\section{Results for the YS and TF models with altruism}

We first investigate the effect of altruism in the YS model.
In order to quantify the inequality in the wealth distribution we
use the Gini index \cite{gindex} defined as
\begin{equation}
G=\frac{\sum_{i=1}^N\sum_{j=1}^N|x_i-x_j|}{2N^2\mu},
\end{equation}
where $\mu$ is the average wealth. A perfect distribution of
wealth where everybody has the same amount of money will
give a value of $G=0$. The other extreme where one individual
has all the money has a Gini value of 1.

It is known that in this model we have condensation: all the money 
ends up in the hands of a single trader, which represents the extreme
case of wealth inequality.  Altruism does not change
this situation. In figure 1 we see the results of several
simulations with 1000 traders that start with an initial
fortune of 100 (these values will remain fixed for all
the simulations in this paper). In the curve with open circles we 
have $p=r=0$, that is, the pure YS model without altruism at all. 
Condensation takes place at about 1000 MCS. If we introduce
altruism, condensation still takes place, the only difference is that
it takes longer to reach. The curve with solid circles has
values of $p=r=0.95$, almost everybody is near totally altruistic,
however, only at the beginning we see a difference in the Gini
index compared to the pure YS model. As time goes by, we quickly
arrive at the condensate phase. Only when we set $p=1$, or everybody
is altruistic, we get saturation of the Gini index. In the figure,
the curve with crosses has $p=1$ and $r=0.1$ which gives a value
of $G=0.62$. This saturation is, however, uninteresting since
the addition of a single non--altruist takes the system to the
condensate phase.

In the TF model condensation does not take place. The effect
of altruism has been studied by using several values for
the fraction and rate of altruism in the model. For each
set of values of $p$ and $r$ we let the system reach a 
stable distribution at about 300 MCS and obtain a value for the
Gini index. As figure 2 shows, at low values of $p$ and $r$
the Gini index is high, resulting in an uneven distribution
of wealth. Higher values of altruism result in a lower
value of $G$, as expected.

\section{Introducing bargaining efficiency in the transactions}

We now introduce the bargaining efficiency concept in the
models. Figures 3 and 4 show the results for the TF and
YS models respectively. It is interesting that the condensation that
occurs in the YS model disappears with the implementation
of this scheme. This is illustrated in figure 4. Note
that, comparing with the TF model (figure 2),
the YS model with bargaining efficiency yields lower
values of the Gini index for the same degree of altruism,
that is, in order to attain a certain value of $G$, we
need lower values of $r$ and $p$ in the YS model with
bargaining efficiency (with $\beta=1$) than in the stand--alone TF model.

If we compare the results for the TF model with and without
bargaining efficiency (figures 2 and 3), we see that $G$
is smaller in the bargaining efficiency case, but only for
small values of the altruism parameters. In fact, for some values of these
parameters, the wealth is better distributed in the stand--alone
TF model. This is shown in figure 5. What the data in this 
figure says is: if you take a TF economy without altruism, 
the addition of bargaining efficiency (giving the poor more chances 
to win) reduces the Gini index, that is, the wealth is more
evenly distributed. However, if, in addition to the fact that
the chances to win are biased in favor of the poor you also
have altruism in your economy, then at a certain point, the pure
TF economy performs better in terms of wealth distribution.

This behavior can be understood in the following way. Take
the case of no altruists at all. In this situation the money
is changing hands all the time, and at any point in time
you can find extremely rich agents and very poor ones, which gives you
a high value for $G$. If in these conditions you give
the poor more chances to win then you are leveling the field
and $G$ diminishes. This is the behavior at low values
of $r$ in figure 5. Now take the other extreme, almost
everyone is altruistic at a high rate. Every time a rich
wins a poor, he will give the poor money so that their
fortunes will be almost the same. This situation gives
you a low value for $G$, but if now you give the poor
more chances to win in addition to the altruism which 
is already helping him, then they benefit in excess
and $G$ increases.

We finally perform a set of simulations to study the
effect of changing the aggressiveness of bargaining,
which is controlled by the parameter $\beta$. The higher
value of this parameter, the most chances has the poorer
of the two traders to win the transaction. These simulations
emphasize the behavior discussed above. By enhancing the bargaining
efficiency, the Gini index decreases, but only when the
altruism is low, for example when $r=p=0.4$ (see the upper
curve in figure 6). When altruism is high the behavior is
interesting, since the Gini index first begins to decrease
when $\beta$ increases, and then reaches a minimum value
and starts to increase for higher values of $\beta$. This 
means that there is an optimum value for the bargaining 
parameter $\beta$ for which the wealth distribution reaches
its more equitable form, at least under the Gini criteria.
In figure 7 we present similar curves as in figure 6 but
for the YS model. In this case the Gini index decreases
monotonically as $\beta$ increases, except for the bottom
curve where there is a very small increment in $G$ after the initial
decrease.

\section{Conclusions}

We have investigated the effect of altruistic behavior in the
YS and TF models with and without bargaining efficiency. We found
that it is no easy to get rid of the condensate phase (when
a single agent owns all the wealth) in the YS model. Only in the
extreme case of 100\% altruists condensation does not take place.
When bargaining efficiency is introduced in the YS model, 
condensation is effectively avoided and a stable wealth distribution
is achieved. The distribution of wealth becomes more equitable
as the altruism is increased. In the stand--alone TF model,
it is also observed that $G$ decreases when altruism is increased.

The introduction of bargaining efficiency gives interesting
results since, for small values of altruism, it has the effect
of decreasing the Gini index and thus leads to a better wealth
distribution, however, at high values of altruism it can have
the contrary effect and increase the value of $G$, and this
behavior is more pronounced in the TF model. This implies
that in these models, when high rates of altruism are present,
there is no necessity of giving the poor more chances to win,
because the wealth distribution will get worst.

An important point is that, despite the fact that we do observe
a better distribution of wealth in both models when the 
altruism is increased, this effect is observed only at
too high values of the altruism parameters. For example, in the
stand--alone TF model, the value of the Gini index without altruism
is about 0.65. From figure 2 we can see that in order to
decrease it only to 0.55 we need to have approximately half
of the population behaving as altruists, with an altruism rate
of about 0.3. A value of $r=0.3$ means that the rich agent will
give the poor one 30\% of the difference in their fortunes, 
a value that is hard to expect in real life. Of course, we are
dealing here with oversimplified models but, as other authors
have found, they can be valuable to shed some light in a very
complex issue, and our findings indicate that altruism cannot
be expected to change the way wealth is distributed in a 
significant way.

{\em Acknowledgments}. We want to thank JL Gonz\'alez-Velarde
for his useful comments.

\newpage\eject

FIGURE CAPTIONS

Figure 1. Time evolution of the Gini index in the YS model. 
Open circles are for a simulation without altruism at all. 
Solid circles are for 95\% of altruists and 0.95 of altruism. 
Crosses are for 100\% of altruists and 0.1 of altruism.

Figure 2. Contour plot of the Gini index as function of the
fraction of altruists $p$ and the rate of altruism $r$ in the TF
model. Results are averaged over at least 5000 realizations.

Figure 3. Contour plot of the Gini index as function of the
fraction of altruists $p$ and the rate of altruism $r$
for the TF model with bargaining efficiency with $\beta=1$. 
Results are averaged over at least 5000 realizations.

Figure 4. Same as figure 3 but for the YS model with
bargaining efficiency, and $\beta=1$.
Results are averaged over at least 5000 realizations.

Figure 5. Curves of $G$ as function of $r$ for a fixed
value of $p=0.9$. The curve with open circles is the
stand--alone TF model, while the filled circles curve
is for the TF model with bargaining eficiency and $\beta=1$.

Figure 6. The Gini index as function of the bargaining parameter
$\beta$ under the TF dynamics. Each curve is for different
altruism parameters, which, from top to bottom curves, are
the following: $r=p=\{0.4,0.5,0.6,0.7,0.8,0.9\}$. Results are
averaged over 1000 realizations.

Figure 7. The Gini index as function of the bargaining parameter
$\beta$ under the YS dynamics. Each curve is for different
altruism parameters, which, from top to bottom curves, are
the following: $r=p=\{0.4,0.6,0.8,0.95\}$. Results are
averaged over at least 1000 realizations.

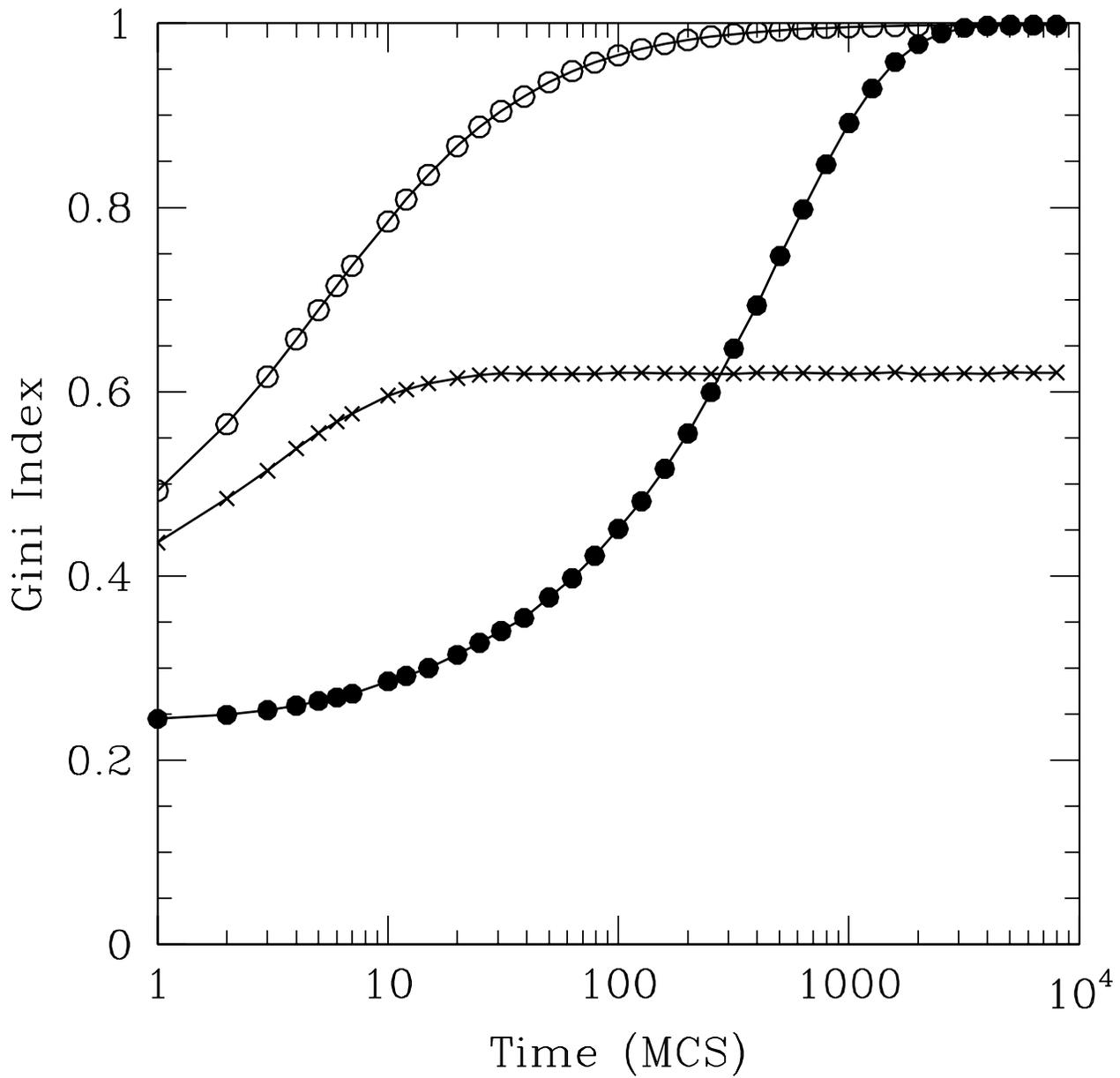
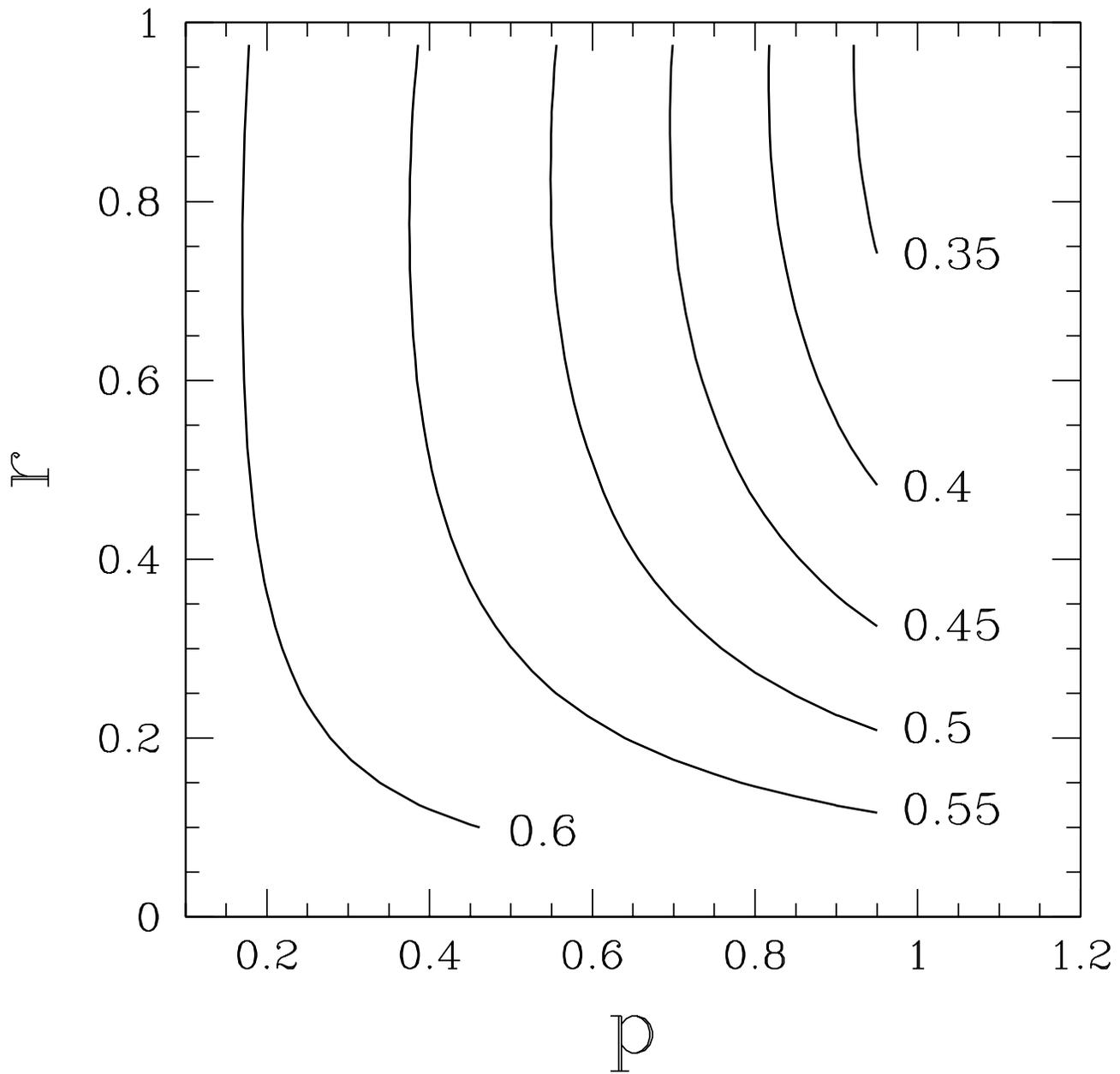
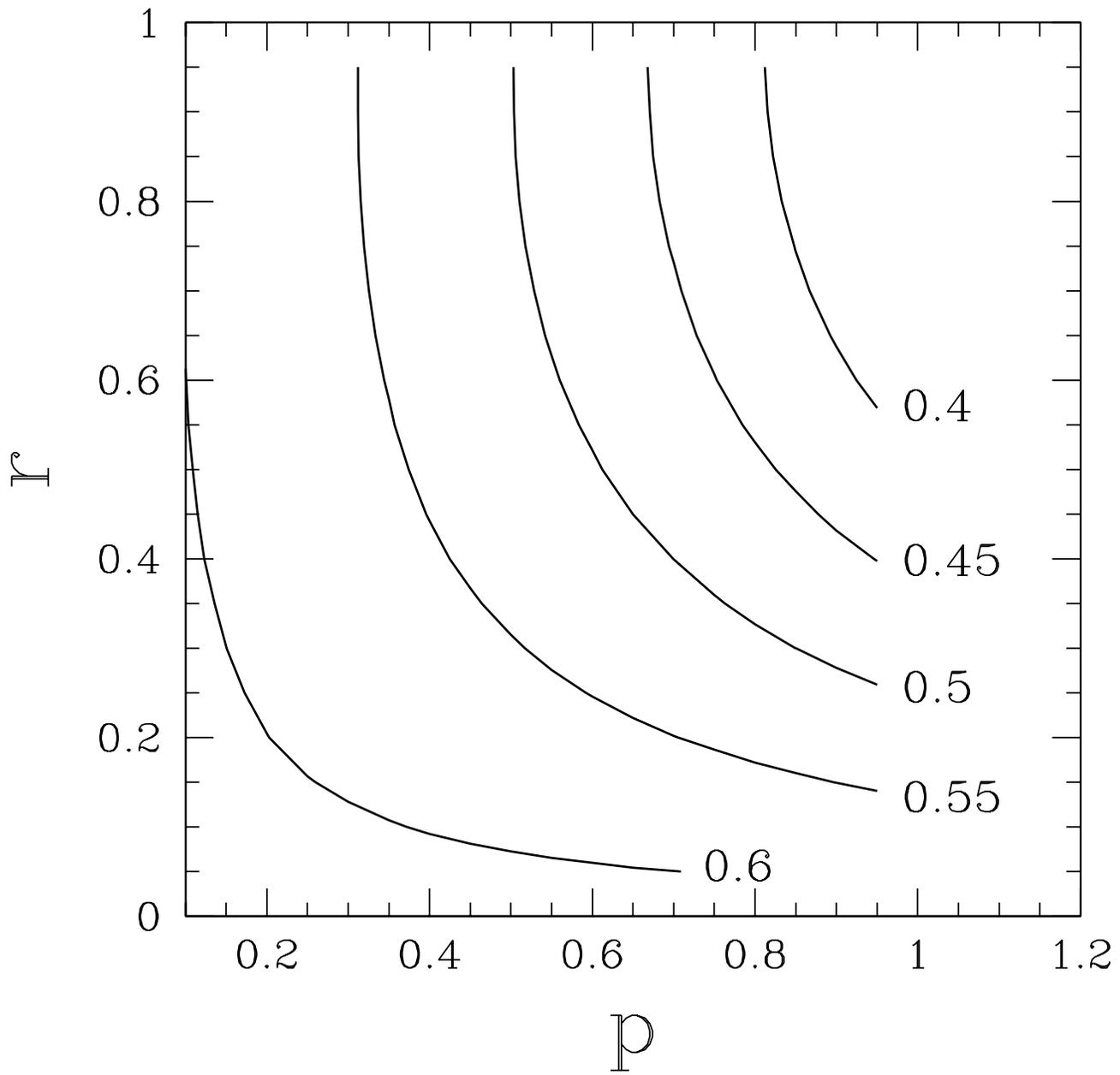
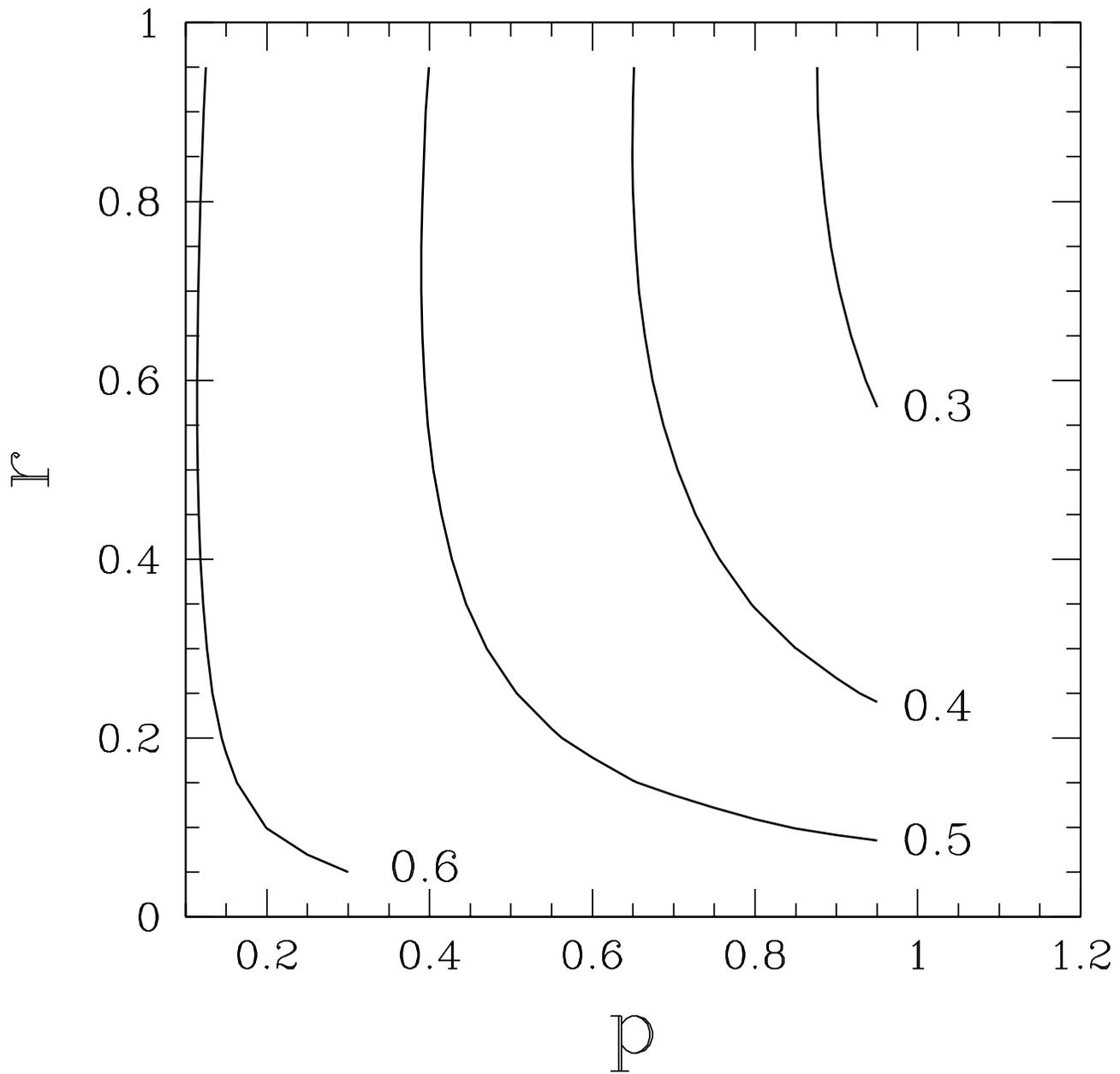
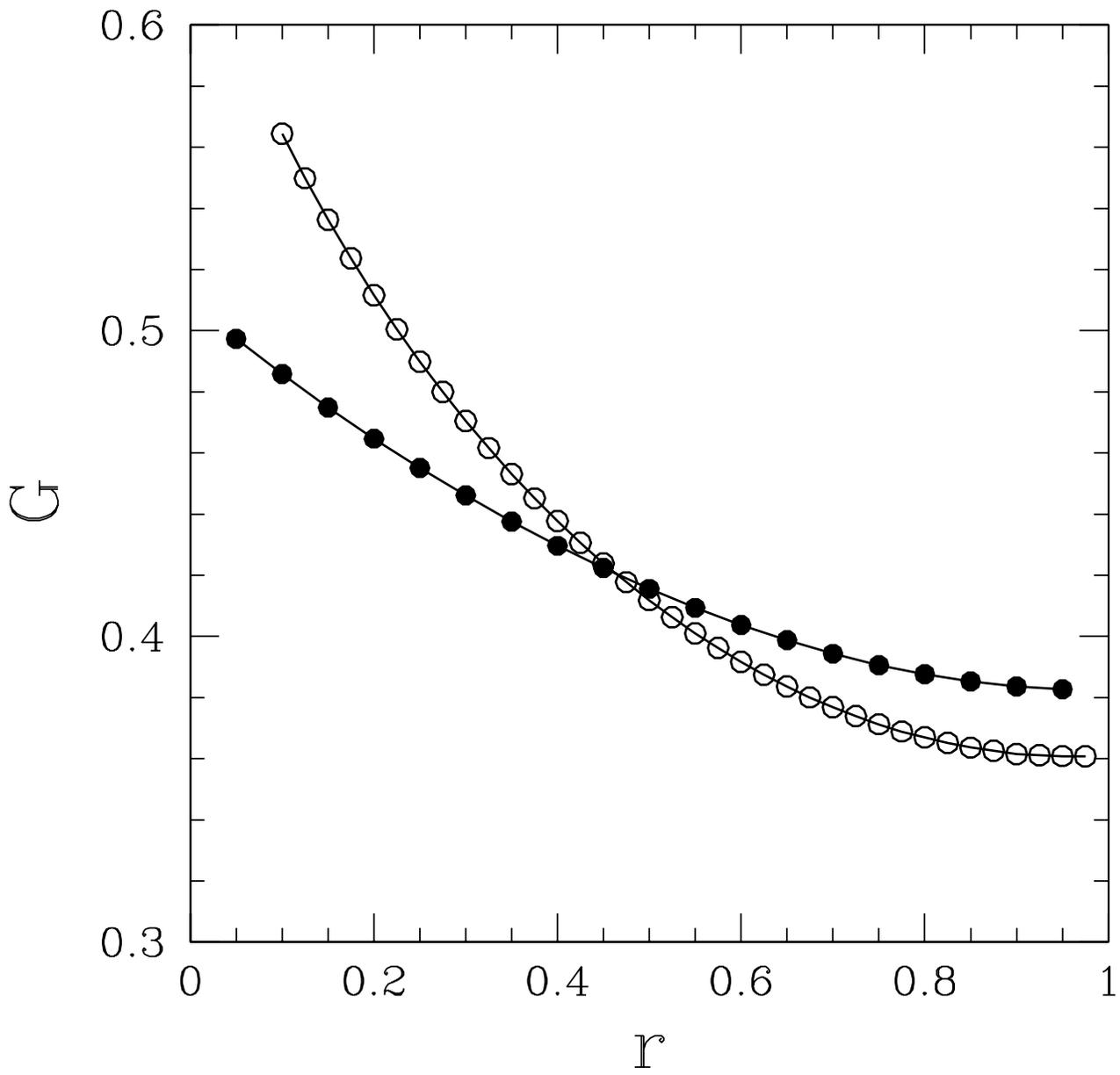
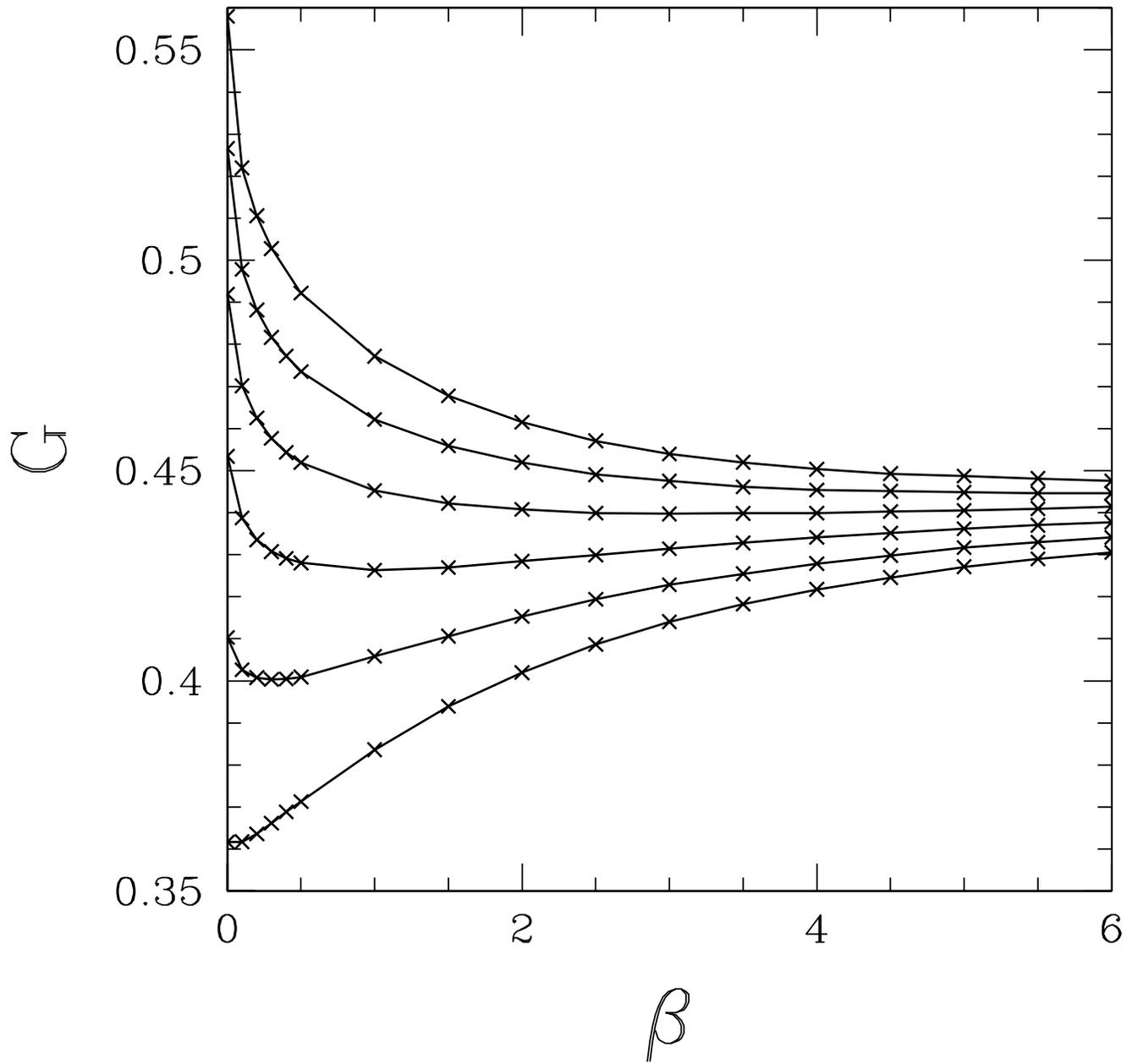
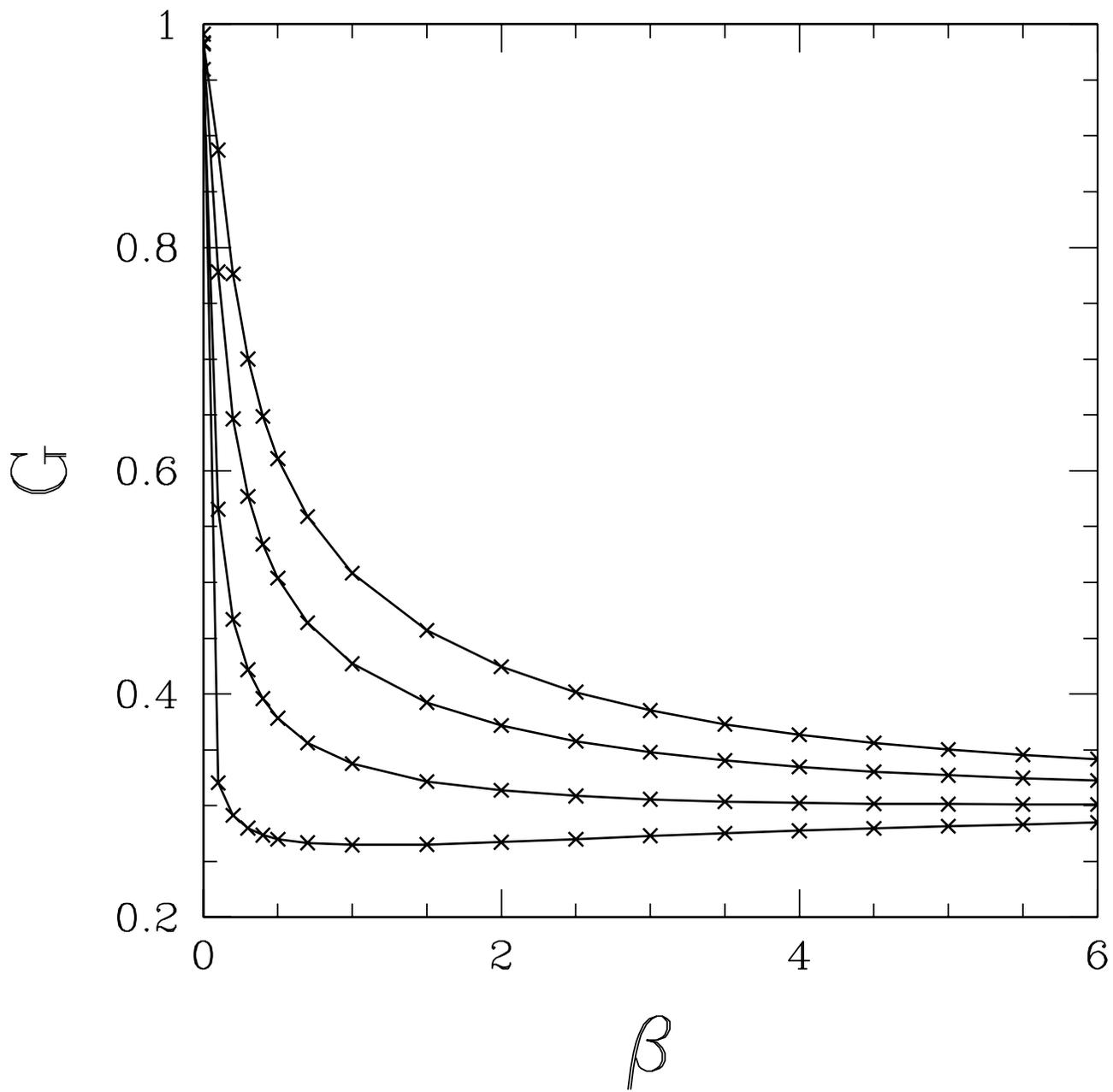
\end{document}